\documentstyle[12pt,epsfig,rotating]{article}  
\newlength{\dinwidth}                       
\newlength{\dinmargin}                      
\def\lsim{\mathrel{\rlap{\lower4pt\hbox{\hskip1pt$\sim$}}
    \raise1pt\hbox{$<$}}}                
\def\gsim{\mathrel{\rlap{\lower4pt\hbox{\hskip1pt$\sim$}}
    \raise1pt\hbox{$>$}}}                
\def\Journal#1#2#3#4{{#1} {\bf #2}, #3 (#4)}

\def\NPB{{\em Nucl. Phys.} B}
 
\def\PLB{{\em Phys. Lett.}  B}

\def\ZPC{{\em Z. Phys.} C}

\begin{document}
\begin{flushleft}


{\tt DESY 96-197    \hfill    ISSN 0418-nnnn} \\
{\tt September 1996}                  \\
\end{flushleft}

\vspace*{1cm}
\begin{center}  \begin{Large} \begin{bf}
Prospects for Measuring $\Delta G$ from Jets at HERA with Polarized Protons
and Electrons\\
  \end{bf}  \end{Large}
  \vspace*{5mm}
  \begin{large}
A. De Roeck$^a$, J. Feltesse$^{b}$, F. Kunne$^b$, M. Maul$^c$,\\ E. Mirkes$^d$,
G. R\"adel$^e$, A. Sch\"afer$^c$, and C.~Y.~Wu$^c$\\ 
  \end{large}
\end{center}
$^a$ Deutsches~Elektronen-Synchrotron~DESY, 
     Notkestrasse~85,~D-22603~Hamburg,~Germany\\
$^b$ DAPNIA,~CE~Saclay,~F-91191~Gif/Yvette,~France\\ 
$^c$ Inst. f. Theor. Phys., Johann Wolfgang Goethe-Universit\"at,
D-60054 Frankfurt am Main, Germany\\
$^d$ Inst.~f.~Theor.~Teilchenphysik,~Universit\"at~Karlsruhe,~D-76128~Karlsruhe,~Germany\\
$^e$ CERN,~Div.~PPE,~CH-1211~Gen\`eve~23,~Switzerland\\

\begin{quotation}
\noindent
{\bf Abstract:}
 The measurement of the polarized gluon distribution
 function $\Delta G(x)$  from photon-gluon fusion processes in 
 electron-proton deep inelastic scattering producing
 two jets has been investigated. The study is based on the MEPJET
 and PEPSI simulation programs.
 The size of the expected  spin asymmetry
 and corresponding statistical uncertainties
 for a possible measurement
 with polarized beams of electrons and protons  at HERA 
 have been estimated.
 The results show that the asymmetry can reach a few percent, and is not
 washed out by hadronization and higher order processes.

\end{quotation}

\section{Introduction}

After confirmation of the surprising EMC result,
that quarks carry  only a small fraction of the 
nucleon spin,
this subject is   being actively studied by several
fixed target experiments
at CERN, DESY and SLAC~\cite{ro}.
So far only the polarized structure functions
$g_1(x,Q^2)$ and $g_2(x,Q^2)$
have been measured.
These structure functions measure predominately the polarized
quark distribution functions, which, as usual,
contain a scheme dependent gluon admixture.
A specific property of polarized structure functions is that this
admixture can be rather large, due to a rather large polarized 
gluon distribution function $\Delta G (x_g)$, as suggested by the 
fact that $\alpha_S (Q^2) \int dx_g \Delta G( x_g,Q^2)$ is renormalization 
group invariant.

The direct measurement of the polarized gluon distribution 
$\Delta G(x_g,Q^2)$
has become the key experiment
in order to understand the 
QCD properties of the spin of the nucleon.
For a collider with polarized electrons and protons with beam energies 
such as for HERA the 
measurement of dijet events offers such a possibility\cite{feltesse}.

\begin{figure}
\mbox{\epsfig{figure=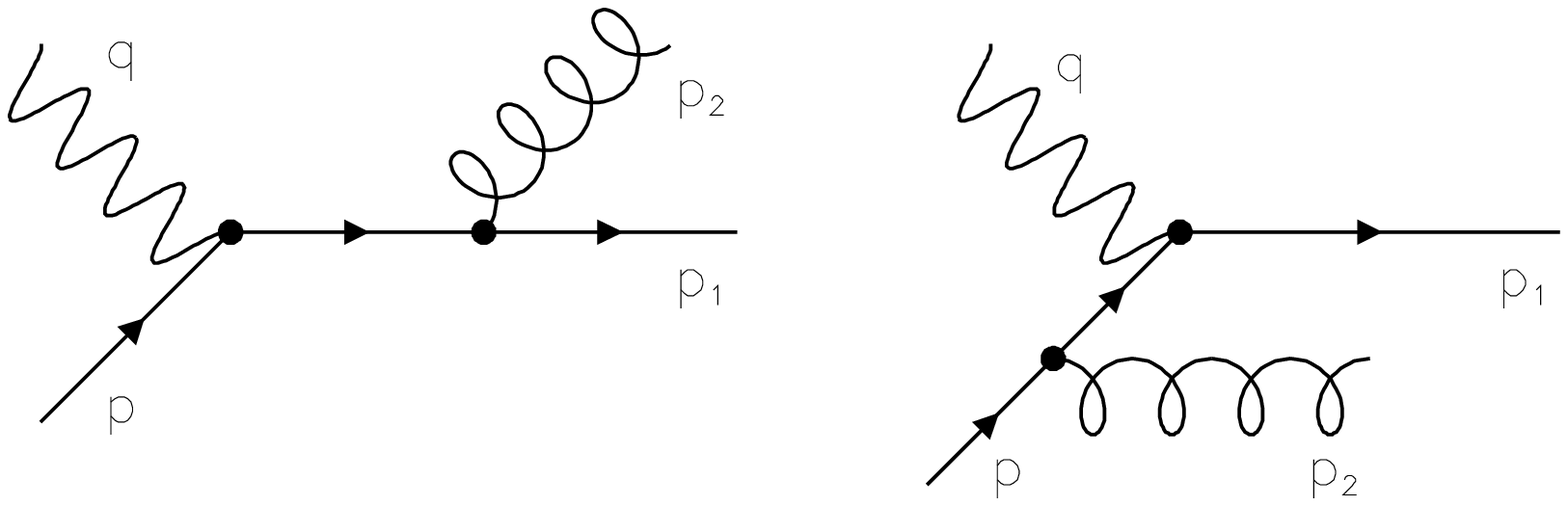,bbllx=0pt,bblly=0pt,bburx=500pt,bbury=200pt,width=8cm}}
\mbox{\epsfig{figure=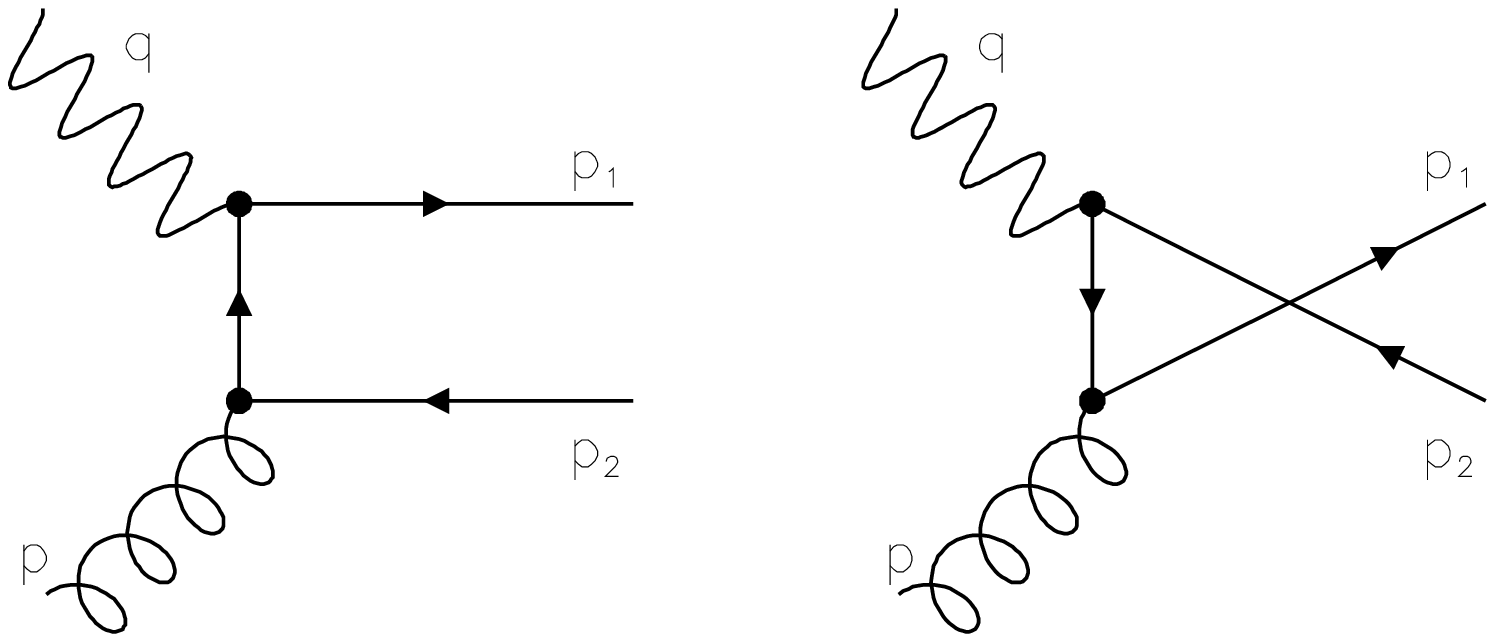,bbllx=0pt,bblly=0pt,bburx=500pt,bbury=200pt,width=8cm}}
\caption{Feynman diagrams for the dijet cross section at LO.}
\label{fig:feyn}
\end{figure}

The gluon distribution  enters at leading order (LO)
in the two-jets production cross section\footnote{In the following 
the jet due to the beam remnant is not included in the number of jets.} 
in deep inelastic scattering (DIS) (see Fig.~\ref{fig:feyn}),
and the unpolarized gluon distribution $G(x_g,Q^2)$
has indeed already been extracted from two-jets events
by the H1 collaboration at HERA.
With a  modest integrated luminosity of  0.24 pb$^{-1}$
collected in 1993, first data on
$x_g G(x_g)$ were extracted from  dijets events~\cite{h1}
at LO,  in  a wide  $x_g$ range $0.002 < x_g < 0.2$,
at  a mean $Q^2$ of $30 ~$ GeV$^2$.
These results  were found to be 
in good agreement with the gluon distribution
extracted at LO from scaling violations of the structure function
$F_2$.
Presently this method is being 
extended to NLO, and first preliminary results were
shown in \cite{rome1,rome2}.
An initial  study of the feasibility of this measurement 
for $\Delta G$ at HERA (mainly
for a  polarized fixed target experiment) was presented in \cite{vogelsang}.

\section{Jet cross sections  in DIS  }
Deep inelastic electron-proton scattering with several partons
in the final state,
\begin{equation}
e^-(l) + p(P) \rightarrow  e^-(l^\prime)+
\mbox{remnant}(p_r) +
\mbox{parton} \,\,1 (p_1) +
\ldots
+\mbox{parton}\,\, n (p_n)
\label{eq1}
\end{equation}
proceeds via the exchange of an
intermediate vector boson $V=\gamma^*, Z$. $Z$-exchange and $\gamma^*/Z$
interference become only important at large $Q^2$ ($> 1000 $ GeV$^2$)
and are neglected 
in the following.
 We denote the 
momentum of the virtual photon, $\gamma^{\ast}$, by $q=l-l^\prime$, 
(minus) its absolute square by $Q^2$,
and use the standard scaling variables Bjorken-$x$
$x_{Bj}={Q^2}/({2P\cdot q})$ and inelasticity $y={P\cdot q}/{P\cdot l}$.
The general structure of the {\it unpolarized}
 $n$-jet cross section in DIS is given by
\begin{equation}
d\sigma^{had}[n\mbox{-jet}] = \sum_a
\int dx_a \,\,f_a(x_a,\mu_F^2)\,\,\, d\hat{\sigma}^a(p=x_a P, 
\alpha_s(\mu_R^2), \mu_R^2, \mu_F^2)
\label{eq2}
\end{equation}
where the sum runs over incident partons $a=q,\bar{q},g$ which carry 
a fraction $x_a$ of the proton momentum.
$\hat{\sigma}^a$ denotes the partonic cross section from which collinear 
initial state singularities have been factorized out 
(in  next-to-leading order (NLO)) at a scale $\mu_F$ and 
implicitly included in the scale dependent parton densities 
$f_a(x_a,\mu_F^2)$. 
For {\it longitudinally polarized} lepton-hadron scattering, the hadronic ($n$-jet) 
cross section is obtained from Eq.~(\ref{eq2}) by replacing
$(\sigma^{had}, f_a\,\, , \hat{\sigma}^a)\rightarrow
 (\Delta\sigma^{had}, \Delta f_a, \Delta\hat{\sigma}^a)$.
The polarized hadronic cross section is defined by
$\Delta\sigma^{had}\equiv\sigma^{had}_{\uparrow\downarrow}
                   -\sigma^{had}_{\uparrow\uparrow}$, where the
left arrow in the subscript denotes the polarization of the 
incoming lepton with respect to the direction of its momentum.
The right arrow stands for the polarization of the proton parallel 
or anti-parallel to the polarization of the incoming lepton.
The polarized parton distributions are defined by
$\Delta f_a(x_a,\mu_F^2)
\equiv f_{a \uparrow}(x_a,\mu_F^2)-f_{a \downarrow}(x_a,\mu_F^2)$.
Here, $f_{a \uparrow} (f_{a \downarrow})$ denotes 
the probability to find a parton $a$ 
in the longitudinally polarized  
proton whose spin is aligned (anti-aligned) to the proton's spin.
$\Delta\hat{\sigma}^a$ is the corresponding polarized
partonic cross section.

In the Born approximation, the subprocesses  
$\gamma^\ast +q \rightarrow q + g$, 
$\gamma^\ast +\bar q \rightarrow \bar q + g$,  
$\gamma^\ast+g \rightarrow q + \bar{q}$  
contribute to the two-jet cross section (Fig.~\ref{fig:feyn}).
The boson-gluon fusion subprocess
$\gamma^\ast+g \rightarrow q + \bar{q}$  dominates the
two-jet cross section at low $x_{Bj}$ 
for unpolarized protons 
(see below) and allows for a  direct
measurement of the gluon density in the proton.
The full NLO corrections for two-jet production in unpolarized
lepton-hadron scattering  are now available~\cite{mi} 
and implemented in the 
$ep \rightarrow n \rm  -jets$ event generator MEPJET,
which allows to analyze 
arbitrary jet definition schemes and 
general cuts in terms of parton 4-momenta.

First discussions about jet
production in polarized lepton-hadron scattering
can be found in Ref.~\cite{zm}, where the jets were defined in
a modified ``JADE'' scheme. 
However, it was found \cite{mi,rheinsberg} 
that  the theoretical uncertainties 
of the two-jet cross section for the ``JADE'' scheme can be very large
due to higher order effects.
These uncertainties are small for the cone scheme and the following 
results are therefore based on the cone algorithm,
which is defined in the laboratory frame.
In this algorithm  the distance 
$\Delta R=\sqrt{(\Delta\eta)^2+(\Delta\phi)^2}$ between two partons 
decides whether they should be recombined 
into a single jet. Here the variables 
are the pseudo-rapidity $\eta$ and the azimuthal angle $\phi$. We 
recombine partons with $\Delta R<1$.
Furthermore, a cut on the jet transverse momenta of $p_T >5$~GeV in the 
laboratory  
frame and in the Breit frame is imposed.
We employ the one loop (two loop) formula for the strong coupling constant
in a LO (NLO) analyses
with 
a value for ${\Lambda_{\overline{MS}}^{(4)}}$
consistent with the value from the parton distribution functions.
In addition a minimal set of general  kinematical cuts
is imposed on the virtual photon and on the final state electron and jets.
If not stated otherwise, we require
5~GeV$^2<Q^2<2500$ GeV$^2$,
$0.3 < y < 1$, an energy cut of $E(e^\prime)>5$~GeV on the scattered 
electron, and a cut on the pseudo-rapidity $\eta=-\ln\tan(\theta/2)$
of the scattered lepton (jets) of $|\eta|<3.5$ ($|\eta|<2.8$).
These cuts are compatible with the existing
detectors H1 and ZEUS, and slightly 
extend the cuts of the H1 gluon analysis from jets.

Let us briefly discuss the choice of the renormalization and factorization
scales $\mu_R$ and $\mu_F$ in Eq.~(\ref{eq2}).
Both the 
renormalization and the factorization scales are tied to the sum of 
parton $k_T$'s in the Breit frame,
\begin{equation}
\mu_R = \mu_F = {1\over 2} \sum_i k_T^B(i) \; .
\end{equation}
Here $(k_T^B(i))^2=2E_i^2(1-\cos\theta_{ip})$, and $\theta_{ip}$ is the 
angle between the parton and proton direction in the Breit 
frame. $\sum_i k_T^B(i)$ interpolates between $Q$, the photon virtuality,
in the naive parton model limit and the sum of jet transverse momenta when
$Q$ becomes negligible, and thus it constitutes a natural scale for jet
production in DIS~\cite{rheinsberg}.


Let us first discuss some results for unpolarized dijet cross sections.
If not stated otherwise,
the lepton and hadron beam energies are 27.5 and 820 GeV, respectively.
With the previous parameters and  GRV parton densities~\cite{grv}
one obtains a LO (NLO) two-jet cross section $\sigma^{had}(2\mbox{-jet})$
of 1515 pb (1470 pb).
Thus the higher order corrections are small. This is  essentially due
to the relatively large cuts on the transverse momenta of the jets.
As mentioned before, the boson-gluon fusion subprocess 
dominates the cross section and contributes  80\% to the LO cross section.

In order to investigate 
the feasibility of the parton density determination,
Fig.~2a  shows the Bjorken $x_{Bj}$ distribution of the
unpolarized two-jet exclusive cross section.
The gluon initiated subprocess clearly dominates the Compton process
for small $x_{Bj}$ in the LO predictions. 
The effective $K$-factor is close to unity
for the total exclusive dijet cross section which is the result of
compensating effects in the low $x$ ($K>$ 1) and high $x$ ($K<$1) regime.

For the isolation of parton distributions we are interested in the 
fractional momentum $x_a$ of the incoming parton $a$ ($a=q,g$, denoted 
$p$ in Fig.~1).  For events
with 
dijet production $x_{Bj}$ and  $x_a$ differ substantially. 
For two-jet exclusive events  
the two are related by
%
$
x_a = x_{Bj} \,\left(1+\frac{{s_{ij}}}{Q^2}\right).
$
%
%
where ${s_{ij}}$ is the invariant mass squared of the produced dijet system.
The $s_{ij}$ distribution for the kinematical region under study 
is shown in Fig.~2b. It is found to  exhibit rather large NLO
corrections as well. The invariant mass squared of the two jets is
larger at NLO than at LO (the mean value of $s_{ij}$ rises to 620~GeV$^2$ 
at NLO from 500~GeV$^2$ at LO).

The NLO corrections to the $x_{Bj}$ and $s_{ij}$ distributions
have a compensating effect on the $x_a$ distribution shown in
Fig.~2c: the NLO and LO predictions have a similar shape.
At LO a direct determination  of the gluon
density is possible from this distribution, after subtraction of the 
calculated Compton subprocess. This simple picture is modified in NLO,
however, and the effects of Altarelli-Parisi splitting and low $p_T$ partons
need to be taken into account more carefully to determine the structure
functions at a well defined factorization scale $\mu_F$ in NLO.

In the following we discuss some results for polarized dijet production.
Our standard set of polarized parton distributions is ``gluon, set A''
of Gehrmann and Stirling~\cite{gs},
for which $\int_0^1 \Delta G(x)dx = 1.8$ at $Q^2 = 4 $~GeV$^2$.
Using the same kinematical cuts as before, 
the LO polarized dijet cross sections 
$\Delta{\sigma}(2\mbox{-jet})$ are shown in the first column of table
\ref{table1}.
The negative value for the polarized dijet cross section 
(--45 pb) is entirely due
to the cross section of the 
boson-gluon fusion process (--53 pb), which is negative
for $x_{Bj}\lsim 0.025$ 
whereas the contribution from the quark initiated subprocess 
is positive over the whole kinematical range.
Note, however, that the shape of the
$x_g$ distribution in the polarized gluon density
is hardly (or even not at all)  constrained by currently 
available DIS data, in particular for small $x_g$.
Alternative parametrizations of the polarized gluon distributions in
the small $x_g$ region,
which are still consistent with all present data \cite{gs1},
can lead to  very different
polarized cross-sections.
The polarized two-jet cross sections for such
parton distributions \footnote{We thank T.~Gehrmann
for providing us with these parametrizations} 
with
$\int_0^1 \Delta G(x)dx = 2.7$ and
$\int_0^1 \Delta G(x)dx = 0.3$ and
at $Q^2 = 4 $~GeV$^2$ are 
shown in column 2 and 3 in table \ref{table1}, respectively.

\begin{table}[thb]
\vspace{3mm}
\caption{LO polarized dijet cross sections 
for different polarized parton distributions (column 1-3).
The contributions from the gluon and quark initiated subprocesses
are shown in the last two lines.
See text for the  details on the kinematics.
}\label{table1}
\vspace{3mm}
\begin{center}
\begin{tabular}{lccc}
        \hspace{2.0cm}
     &  $\int_0^1 \Delta G(x)dx = 1.8$
     &  $\int_0^1 \Delta G(x)dx = 2.7$
     &  $\int_0^1 \Delta G(x)dx = 0.3$\\[2mm]
\hline\\[-2mm]
$\Delta\sigma\mbox{2-jet}$    & $-45$   pb & $ -67.5$  pb  & $-3$   pb   \\
$\Delta\sigma^g\mbox{2-jet}$  &  $-53$  pb &  $-76$   pb  & $-10$  pb    \\
$\Delta\sigma^q\mbox{2-jet}$  &   8   pb &  8.5   pb  &  7   pb    \\[2mm]
\hline\\
\end{tabular}
\end{center}
\end{table}
The fractional momentum distributions $x_a$ 
of the incident parton ($a=q,g$), shown in Figs.~2d-f
for the three sets of
polarized parton densities,
demonstrate the sensitivity of the dijet events to 
the choice of these parametrizations, particularly
in the lower $x_a$ range.
Note, that the fractional momentum distributions 
are again related to $x_{Bj}$
by
$
x_a = x_{Bj} \,\left(1+\frac{{s_{ij}}}{Q^2}\right).
$
The corresponding $x_{Bj}$ and $s_{ij}$ distributions 
are not shown here.


%
%
\begin{figure}[htb]
\epsfig{figure=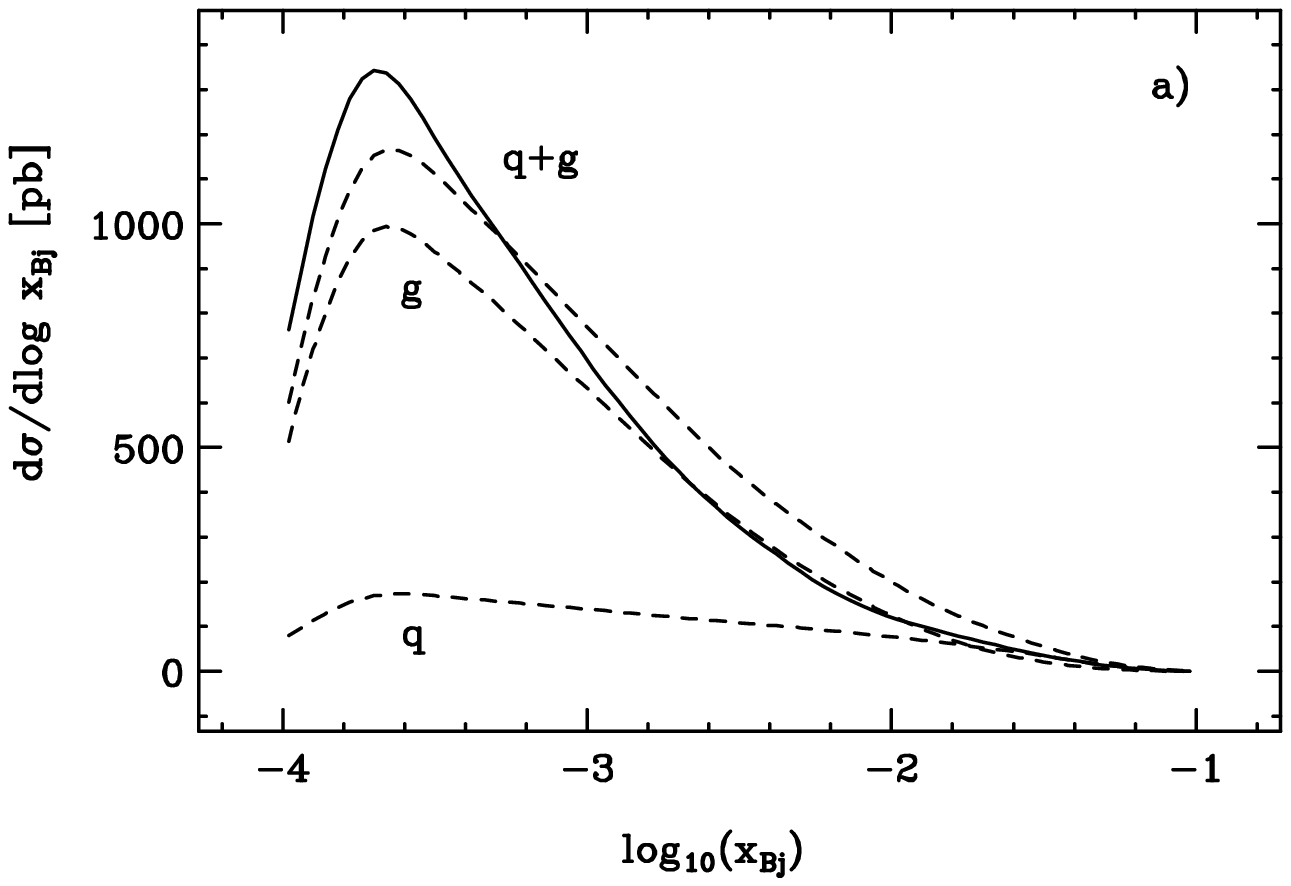,bbllx=30pt,bblly=270pt,bburx=450pt,bbury=570pt,width=8.3cm}
\epsfig{figure=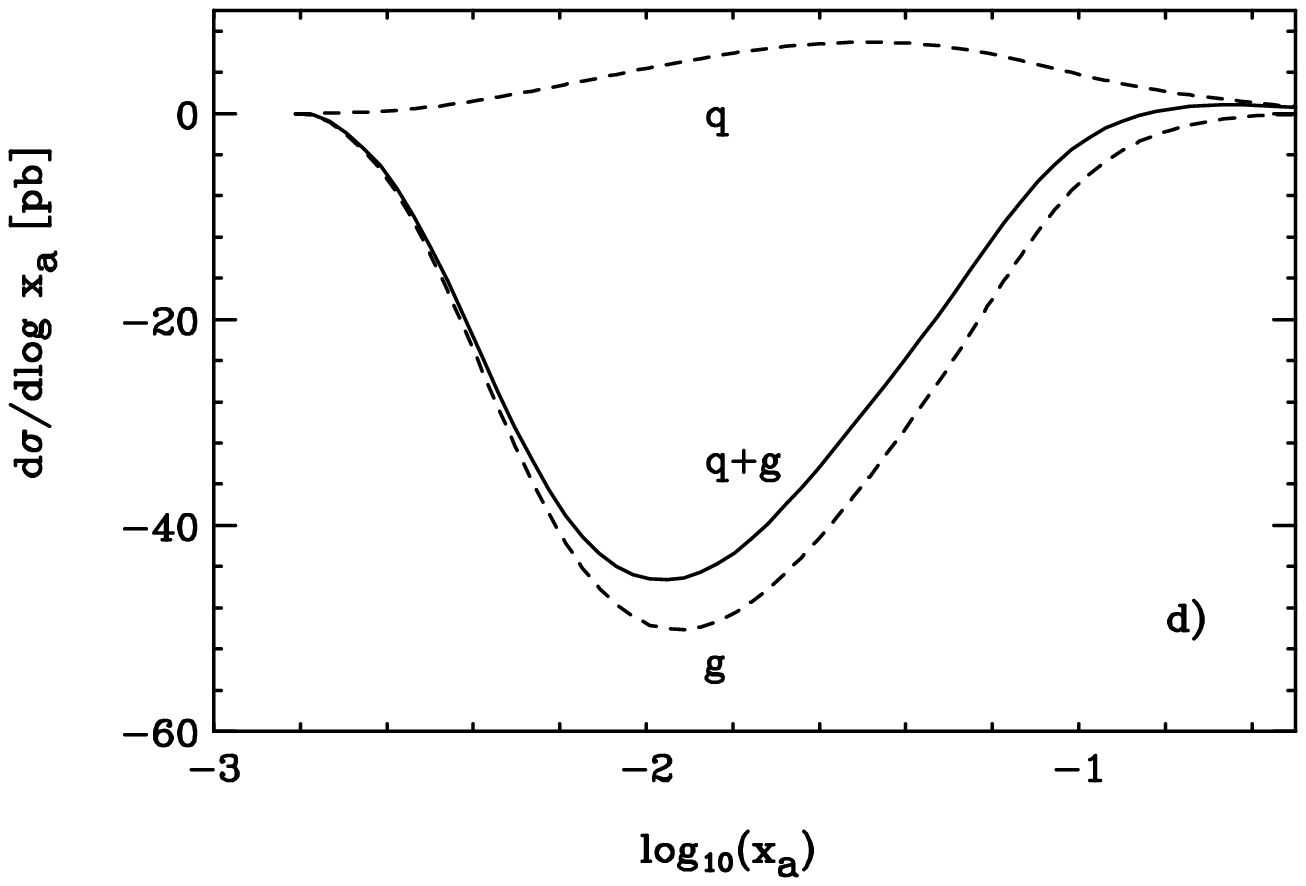,bbllx=60pt,bblly=270pt,bburx=480pt,bbury=570pt,width=8.3cm}
\epsfig{figure=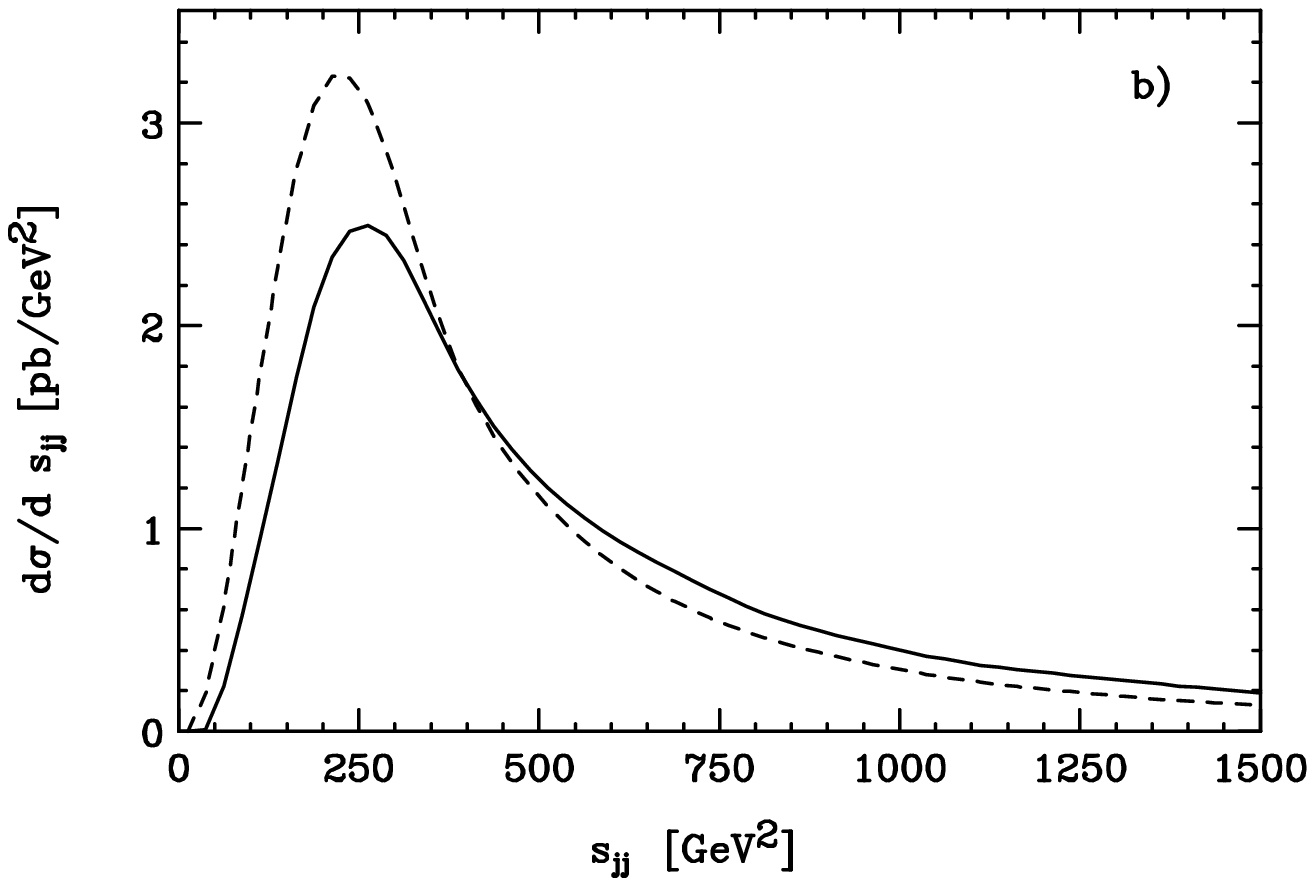,bbllx=30pt,bblly=270pt,bburx=450pt,bbury=570pt,width=8.3cm}
\epsfig{figure=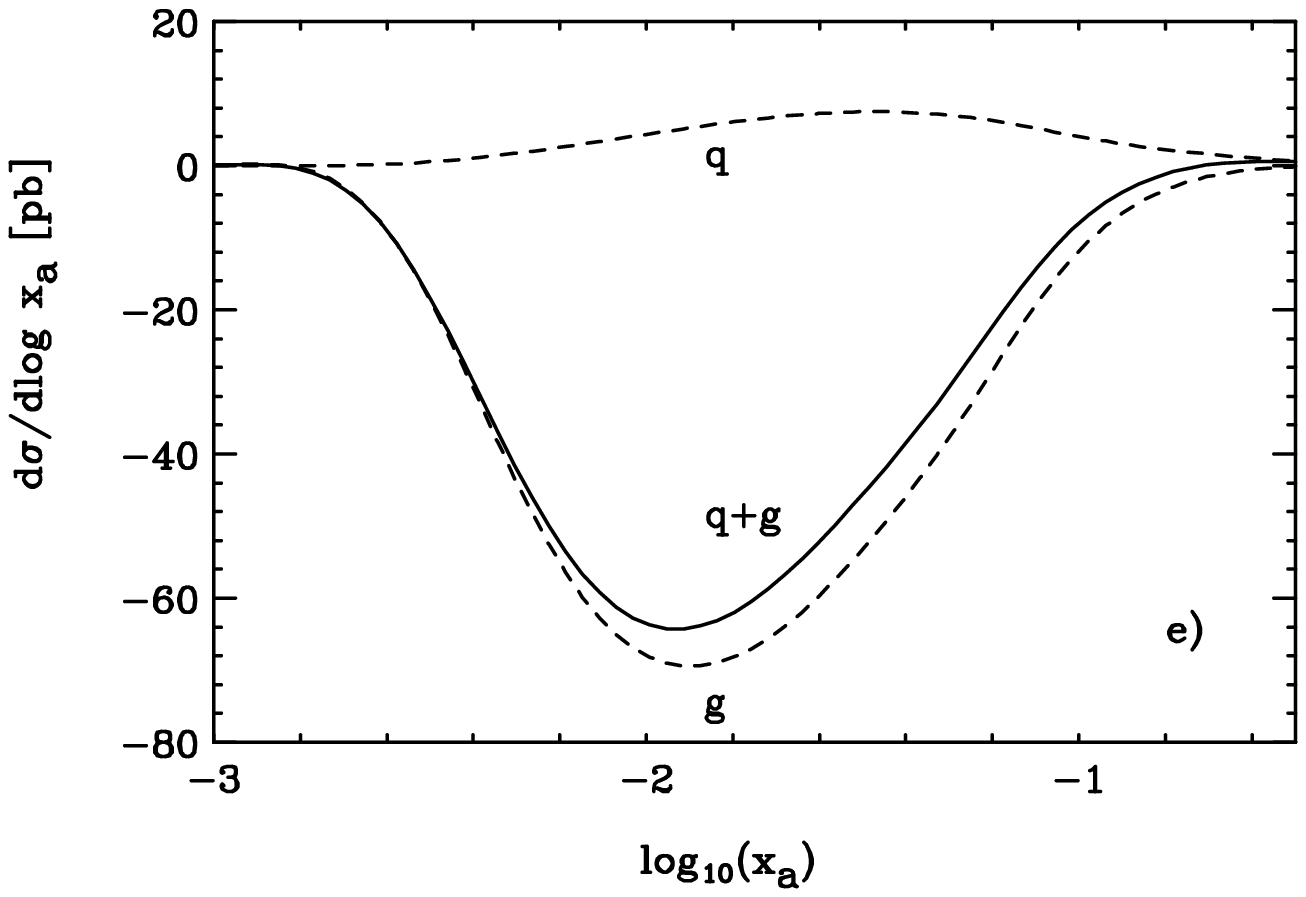,bbllx=60pt,bblly=270pt,bburx=480pt,bbury=570pt,width=8.3cm}
\epsfig{figure=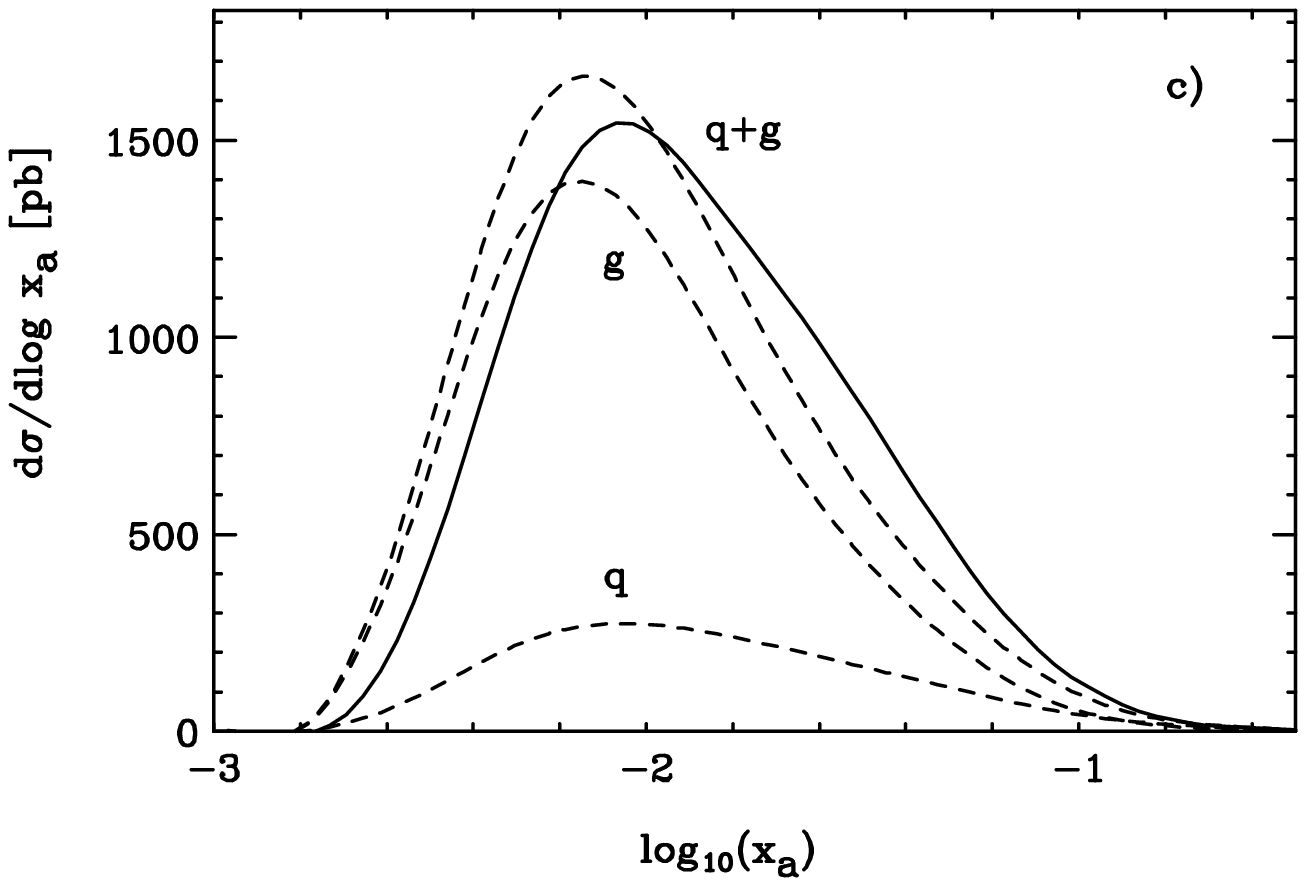,bbllx=30pt,bblly=270pt,bburx=450pt,bbury=570pt,width=8.3cm}
\epsfig{figure=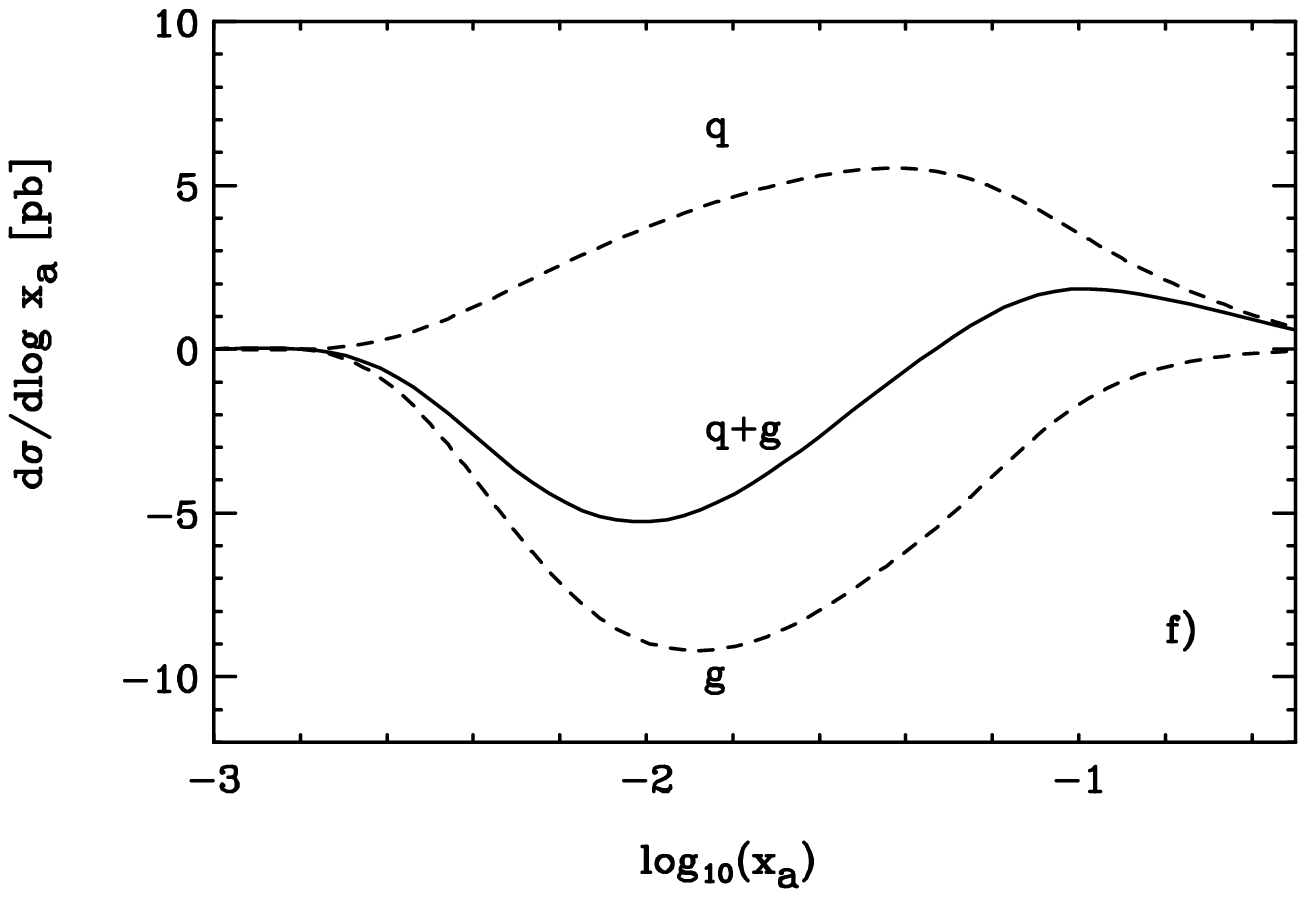,bbllx=50pt,bblly=270pt,bburx=470pt,bbury=570pt,width=8.3cm}
\caption{a) Dependence of the  {\it unpolarized}
two-jet cross section
on Bjorken $x_{Bj}$ for the
quark and gluon initiated subprocesses and for the 
sum. Both LO (dashed) and NLO (solid) results are shown;\,\,
b) Dijet invariant mass distribution in LO (dashed) and in NLO (solid)
for unpolarized dijet production;\,\,
c) Same as a) for the $x_a$ distribution, 
$x_a$ representing the 
momentum fraction of the incident parton at LO;\,\,
d)  Dependence of the LO {\it polarized}
two-jet cross section
on  $x_a$ for the
quark and gluon initiated subprocesses (dashed) and for the 
sum (solid). Results are shown for the polarized parton distributions
``gluon, set A'' of Gehrmann and Stirling~\protect\cite{gs},
for which $\int_0^1 \Delta G(x)dx = 1.8$ at $Q^2 = 4 $~GeV$^2$;\,\,
e) same as d) for $\int_0^1 \Delta G(x)dx = 2.7$ at $Q^2 = 4 $~GeV$^2$;\,\,
f) same as d) for $\int_0^1 \Delta G(x)dx = 0.3$ at $Q^2 = 4 $~GeV$^2$.
}
\label{fig2}
\end{figure}

\section{Experimental asymmetries}

\subsection{Parton level studies}

In order to study the feasibility and 
the sensitivity of the measurement of the
spin asymmetry at HERA, we have assumed
polarizations of 70\%  for both 
the electron and the proton  beams and
statistical errors were calculated for a total luminosity of
200 pb$^{-1}$ (100 pb$^{-1}$ for each polarization).

\begin{figure}
\begin{sideways} \put(-200,30){ \bf $ (\sigma^{\uparrow\downarrow} 
-\sigma^{\uparrow\uparrow})/(\sigma^{\uparrow\downarrow}+
\sigma^{\uparrow\uparrow})$}
\put(-400,30){ \bf $ (\sigma^{\uparrow\downarrow} 
-\sigma^{\uparrow\uparrow})/(\sigma^{\uparrow\downarrow}+
\sigma^{\uparrow\uparrow})$}
\end{sideways}
\psfig{figure=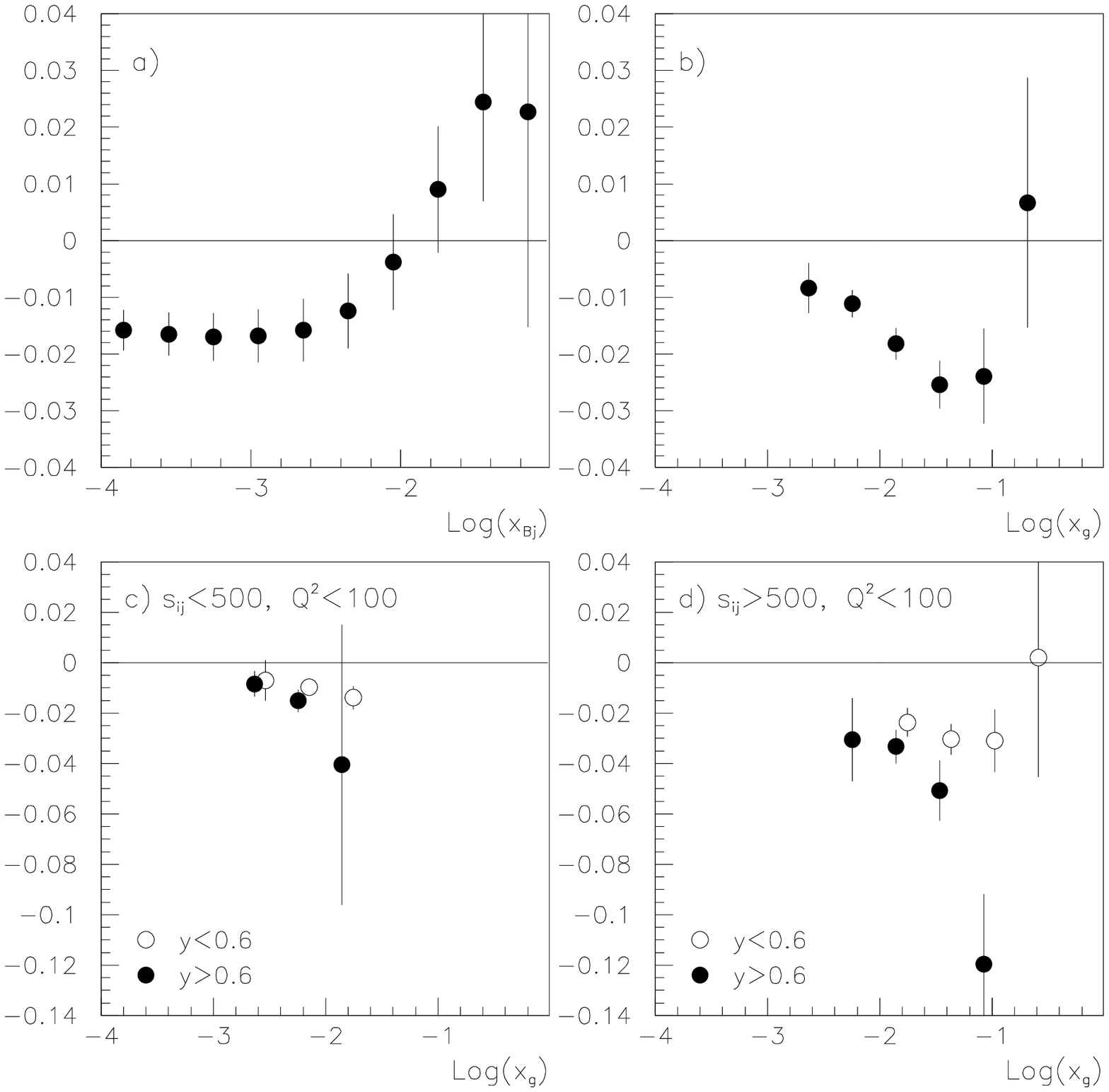,bbllx=-35pt,bblly=0pt,bburx=600pt,bbury=600pt,width=17cm}
\caption{Expected asymmetries as a function of $x_{Bj}$ (a)
and $x_g$ (b-d) for a luminosity of 200 pb$^{-1}$
and beam polarizations $P_e = P_p = 70 \%$;
Fig.a) and b):nominal cuts;
Fig.c) and d): $Q^2 < 100 $~GeV$^2$;
In Fig.c) and d) data are separated in bins of $s_{ij}$ and $y$.
}
\label{fig:asy}
\end{figure}

The expected experimental asymmetry
$<A>
=\frac{\Delta\sigma^{had}(\mbox{2-jet})}{\sigma^{had}(\mbox{2-jet})}$
under these conditions is shown in
Fig.~\ref{fig:asy}a, as a function of $x_{Bj}$ and
in Fig.~\ref{fig:asy}b-d as a function of $x_g$.
Figures a) and b) correspond to the nominal kinematical cuts
defined previously, except for the $Q^2$ range which was extended
to lower values $2 < Q^2 < 2500$~GeV$^2$ and $10^{-5} < x_{Bj} < 1$.
The cross section integrated over all variables is $2140$ pb,
where  82\% of the  contribution 
comes from gluon--initiated events. The
asymmetry averaged over all variables is $<A> = -0.015 ~\pm ~0.0015$.
It is negative at low $x_{Bj}$ and becomes positive
at $x_{Bj}>0.01$ as expected from Fig.~2.

In Figs.~\ref{fig:asy}c,d  a further cut was made:
$Q^2 < 100$~GeV$^2$. 
This cut permits to reject the positive contributions to the asymmetry
coming from high $Q^2$ (equivalent to high $x_{Bj}$) events,
where the contribution of quark--initiated events is higher.
All the remaining events were separated in two
bins in $s_{ij}$ --the invariant mass of the dijet--
and two bins in $y$, as the asymmetry and expected NLO corrections are
very sensitive to these two variables.
Fig.\ref{fig:asy}c
  corresponds to low invariant masses ($s_{ij} < 500$~GeV$^2$),
and Fig.~\ref{fig:asy}d to high ones ($s_{ij} > 500 ~$GeV$^2 $).
Open  points show low  $y$ values ( $y < 0.6$),
and closed points, high  $y$ values ( $y > 0.6$).
In the best case, the asymmetry reaches  values as high as
$12\%$ (Fig.~\ref{fig:asy}d).

Reducing the beam energy to 410 GeV, instead of the nominal
820 GeV, does not improve the signal in average,
although the mean value of $y$ is higher.
The asymmetry signal increases only for a few points around $x_g>0.1$,
since a higher incident energy probes slightly higher values of $x_g$.

\subsection{Hadronization and detector effects}
So far we discussed cross sections only on the parton
 level and to lowest order.
While, on the level of matrix elements, one can only calculate
leading order and next to leading order corrections,
in real experiments one always encounters a coherent superposition of
contributions of all orders.
It is therefore important
to
investigate next how
the picture changes when one takes the effects of higher 
orders, the fragmentation
of the jets into hadrons and   detector smearing into account.
These effects have been studied in two ways: 
 i) a different program 
called PEPSI (see below) was used, which is  a full LO  lepton-nucleon
scattering Monte Carlo program for unpolarized and polarized interactions,
including fragmentation;
ii) parton showers, to emulate the higher orders, and parton
fragmentation have been added to the LO matrix elements of MEPJET.
PEPSI  was further used to  study 
effects of detector
smearing.

The PEPSI program is the polarized extension to the unpolarized
lepton-nucleon Monte Carlo program LEPTO 6.2~\cite{LEPTO}.
It adds to the unpolarized cross section, which 
is a convolution of the partonic cross section $ \sigma_p$ and
the unpolarized structure functions $q(x,Q^2)$, the respective
polarized cross section as a convolution of the polarized
partonic cross section $\Delta \sigma_p$ and the polarized
parton densities $ x \Delta q (x) $:
\begin{equation}
d \sigma \sim
  \int_{x_{p,min}}^{ x_{p,max}} \frac{dx_p}{x_p} q(x_p/x)  d\sigma(x_p,y) 
  +POL \int_{x_{p,min}}^{ x_{p,max}} 
\frac{dx_p}{x_p}\Delta q(x_p/x)  d \Delta \sigma(x_p,y)
\end{equation}
Here $x$ is the Bjorken $x_{Bj}$, while $x_p$
is defined via $x_p = x_{Bj}/x_a$, with $x_a$  the fraction of the
target momentum carried by the initial parton, as defined before.
  $POL$ is a parameter
which is +1 for lepton and nucleon spin being antiparallel to each 
other and $-1$ in the case where the electron and nucleon spins are
parallel to each other. 
\begin{figure}[htb] \centering
\epsfig{file=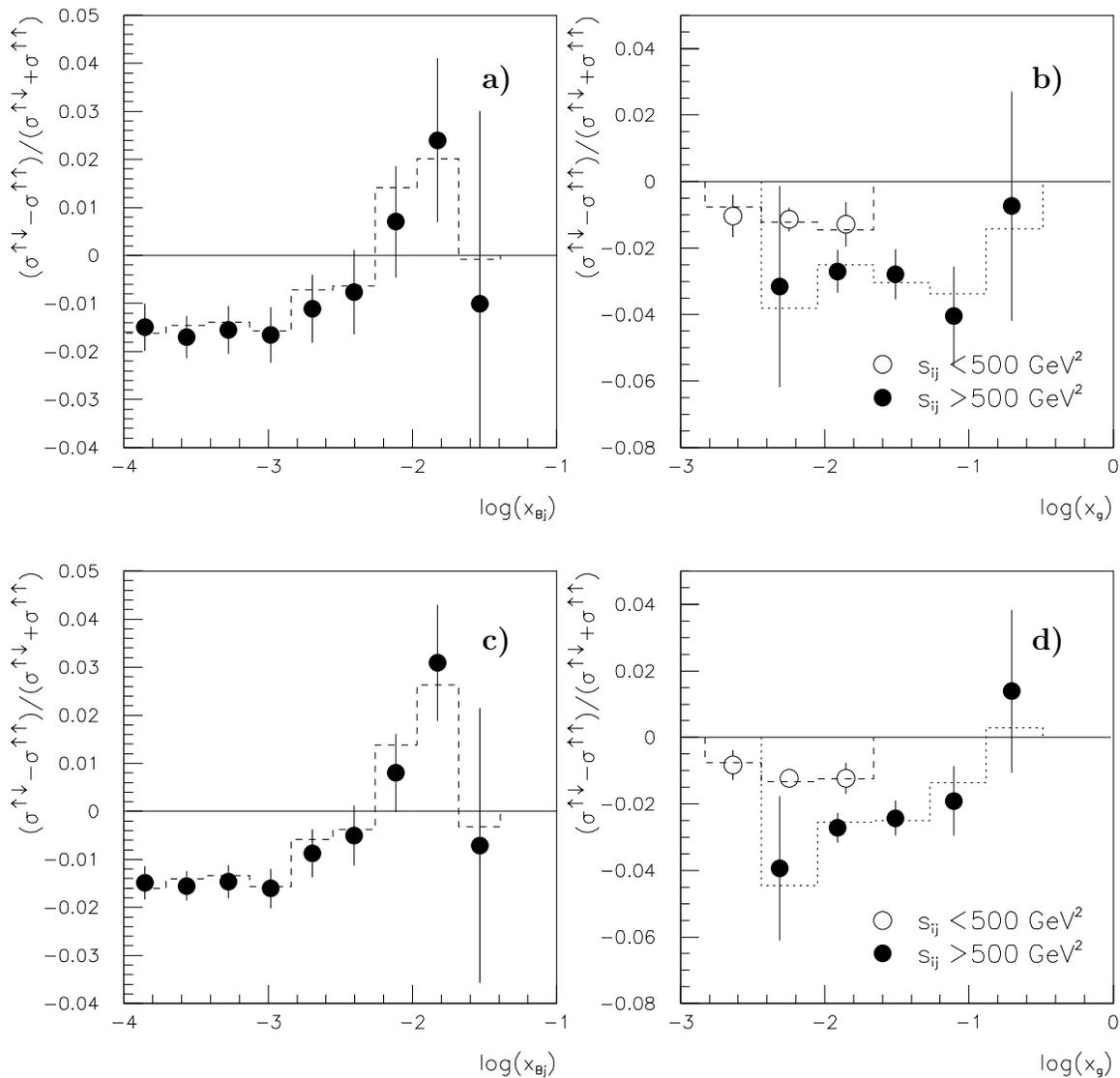,bbllx=0pt,bblly=145pt,bburx=600pt,bbury=670pt,height=16cm}
\begin{picture}(0,0) 
\put(-40,420){ \bf {a)}} 
\put(175,420){ \bf {b)}} 
\put(-40,200){ \bf {c)}} 
\put(175,200){ \bf {d)}} 
\end{picture}
\caption{ Expected asymmetries as a function of $x_{Bj}$ (a,c) and 
$x_g$ (b,d) for a luminosity of 200 pb$^{-1}$ and beam polarizations
$P_e = P_p = 70\%$. The figures (b,d) have an additional cut $Q^2< 100 $
GeV$^2$ on top of the nominal cuts, explained in the text.
Figs.~(a,b) are for the polarized parton set GS-A, Figs.~(c,d) for 
GS-C \protect\cite{gs}.
}
\label{fig:gsa}
\end{figure}
The unpolarized leading order (LO) partonic cross section used already in 
LEPTO is described in \cite{PR80}, the longitudinally polarized
LO cross section is given in \cite{MSV91}. The leading order polarized
cross section contains two subprocesses: the gluon bremsstrahlung and
the photon-gluon fusion, see Fig. \ref{fig:feyn}. 
The partons fragment into hadrons via string fragmentation as implemented
in JETSET\cite{JETSET}.
For the polarized parton density 
functions there are two different sets \cite{gs,GRSV95} implemented 
which both  appertain to the unpolarized parton density 
functions given in set \cite{grv}.  

Event samples corresponding to 
200 pb$^{-1}$ are generated with a similar event selection
as for the partonic analysis in Section 3.1.
The beam polarizations were taken to be
$P_e = P_p = 70\%$.
The kinematic region $5 < Q^2 < 2500$ GeV$^2$, $0.3< y< 0.8$ was selected
(the cut in $y$ corresponds roughly to $E(e') > $ 5 GeV).
 Jets are searched for with the cone algorithm 
(radius 1) with $p_T > 5 $ GeV in the pseudorapidity region
$-2.8 < \eta < 2.8$, conform with for example the present H1 detector. 
The LO 2-jet cross section thus calculated with PEPSI is 
1511 pb, 82\% of which is contributed by the boson-gluon fusion process.
These numbers compare well with those of MEPJET,  given in section 2.

To increase statistics we further study  jets with $p_T > 4$ GeV, and
repeat the studies of section 3.1. This does not change the asymmetries
significantly.
In Figs.~\ref{fig:gsa}a,b the results  for  the asymmetry are shown as
function of $x_{Bj}$ and $x_g$ for similar cuts as in Fig.~\ref{fig:asy},
using the ``gluon set A''. For Fig.~\ref{fig:gsa}b the $Q^2$ range has 
been limited to 100 GeV$^2$.
The calculations are shown at the parton level (dashed and dotted lines)
and at the detector level (open and closed 
points). A calorimetric energy resolution of 
$\sigma_E/E = 0.5/\sqrt{E (\mbox{GeV})}$ is used to simulate the detector 
response.

The first observation is that the asymmetries produced by PEPSI are very 
similar to the ones from MEPJET. Secondly the asymmetry at the detector 
level is well correlated with the asymmetry at the parton level.
This means that present detectors at HERA are well prepared to make 
this measurement. It was verified that this conclusion still holds
with a calorimetric energy 
resolution which is twice worse. Also a miscalibration of
the energy scale of $2\%$, a number within reach at the time 
of this measurement, does not disturb the correlation significantly.

The average asymmetry $<A>$ amounts to $-0.016\pm0.002$ at the parton 
level and $-0.015\pm0.002$ at the detector level.
For the selected kinematic region 40\% of the events accepted on the 
parton level do not enter the detector event sample, and 15\% of the 
final selected events have a parton kinematics outside of the measured
region. When the region is restricted to $s_{ij} > 500$ GeV$^2$ these
migrations are reduced to 25\% and 12\% respectively.

In Fig.~\ref{fig:gsa}c,d the results are shown for  the asymmetry 
for a different set of polarized parton distributions: 
 ``gluon set C''\cite{gs}. 
These exhibit a smaller asymmetry around $x_g \sim 0.1$ than the 
``gluon set A'',
as can be seen from the parton level curves. This difference survives 
after the detector smearing, and thus measurements of this quantity 
can help to discriminate between different sets of parton distributions.

\subsection{Higher order effects}
The studies above were based on a program that includes the LO matrix 
elements and fragmentation, but no higher order QCD effects.
The effect of the latter, and the comparison with the effect
of fragmentation, was studied in a dedicated analysis whereby
 the leading order
MEPJET program used in section 2 and 3.1 was 
supplemented by two packages. The first one
is the program PYTHIA57 \cite{PYTHIA} which  simulates higher order effects
in the framework or initial and final state parton showering and
the second one is the program JETSET74 \cite{JETSET} for hadronization, 
as used in the PEPSI Monte Carlo.

Following a previous publication\cite{mi}, 
the following cuts are used for the kinematic variables:
$ 40 < Q^2 < 2500\; {\rm GeV}^2$,
$ 1 < W^2 < 90000\; {\rm GeV}^2$,
$ 0.001 < x_{Bj} < 1$, and
$ 0.04 < y <1$.
We further impose for the outgoing lepton a cut on the minimal 
energy of 10 GeV and on the rapidity range $|y| < 3.5 $.

Jets are generated 
via leading order 
matrix elements with the MEPJET program
and then fed into PYHTIA  as an external process,
which adds initial and final state parton showering.
Thus all QCD processes are included except for 
 four-gluon vertices. A more detailed description of 
parton showering can be found in \cite{BS87,Sj85}. 
For the showering scale, initial and
final we used both times $Q^2$. We will investigate scale dependence
in a forthcoming publication.
The  parton density functions 
 set MRS D--' \cite{MRS93} was taken.
For the final hadrons and partons we also choose the rapidity cut
$|y| < 3.5$. The jet scheme for detecting hadron
and parton jets is the cone scheme as described before with the
maximal cone distance one.
A sharp cut off of  5 GeV
for the jet transverse momentum was taken.

In the showering process jets with small $p_T$ can
branch into jets with $p_T$ $> 5\; {\rm GeV}$. 
The string fragmentation mechanism contributes in a similar way
thus enhancing the magnitude of the total cross section within 
the imposed detector cuts. 

\begin{figure}
\hspace*{3cm}
\epsfig{figure=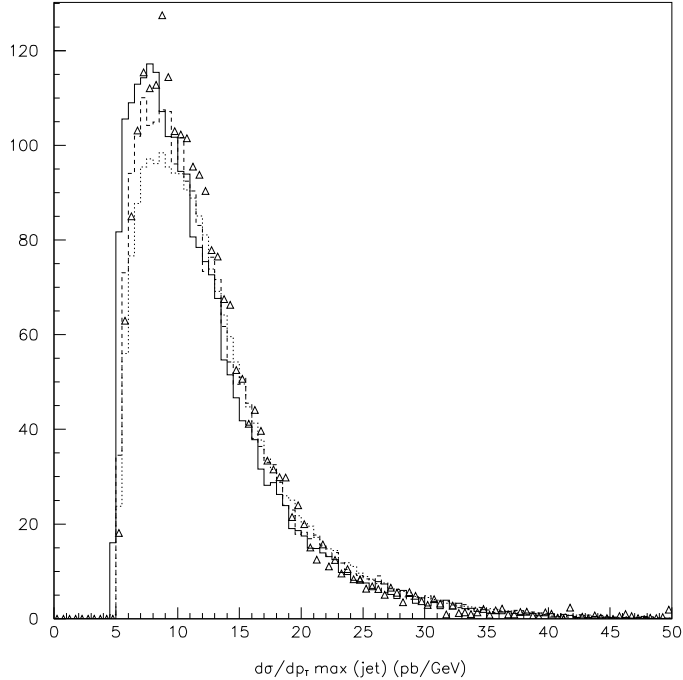,bbllx=2pt,bblly=2pt,bburx=560pt,bbury=560pt,width=10cm,clip=}
\caption{Simulation of higher order and fragmentation effects:
triangles: NLO cross section, solid histogram: LO + parton shower +
fragmentation, dashed histogram: LO + parton shower, dotted histogram: 
only LO matrix elements.}
\label{fig:had}
\end{figure}

Fig.~\ref{fig:had}  shows the unpolarized
inclusive two-jets cross section
for the sum of gluon and quark initiated events. 
The solid histogram shows the differential cross section versus 
the maximum of $p_T$ for the analyzed two-jet events 
after parton showering and hadronization.
It is in magnitude and shape comparable to the NLO order matrix element
calculation  by the  MEPJET program,
represented by the open triangles. A small shift
in the direction of small $p_{T, \rm max}$ is observed,
caused by the fragmentation process.
The dashed histogram shows the cross section for parton showering 
without fragmentation. The absolute cross section is smaller than the
NLO cross section, but the shapes agree well, without any significant shift.
Hence we do not expect that the results calculated with PEPSI, as shown
above, will be
strongly affected by higher order processes. In comparison, 
the dotted histogram gives the leading order 
result from the MEPJET program.

The figure shows that the deviations are significant only for 
$p_{T,\rm  max}$ smaller
than 15 GeV, i.e for soft jets where also next to leading order effects
are expected to be large. For larger $p_{T,\rm max}$ the calculations are
in good agreement.
The differences between the various curves can serve as a 
measure for the systematic errors. It has to be noted that
the effects of showering and fragmentation are large
if and only if the NLO effects are large, too. 
First studies with higher order effects included in PEPSI confirm that
the the asymmetry survives in the presence of parton showers. With the 
present analysis the 
asymmetry however reduces to $<A> = 0.009\pm 0.002$ (parton level)
and the shape of the 
asymmetry distribution as function of $x_g$ changes somewhat with 
$x_g$. Improved methods to determine $x_g$ may restore the original
distributions and are subject of a future study.

\section{Concluding Remarks}
The  results show that if the assumed luminosity
and beam polarizations can be delivered at HERA,
the present detectors H1 and ZEUS will be in 
a comfortable position to measure a  spin asymmetry
of a few per cent in average,  with a
few per mile statistical precision, using two-jet events.
The asymmetries survive at the hadron and detector level.
In order to minimize the experimental systematic uncertainties,
it  is desirable  to have in the HERA ring
bunch trains of protons with alternating helicity.
On the theoretical side, NLO QCD corrections  are needed.
The NLO corrections reduce the renormalization $\mu_R$ and factorization scale
$\mu_F$  dependence 
(due to the initial state collinear factorization)
in the LO calculations and thus reliable predictions
in terms of a well defined strong coupling constant and
scale dependent parton distributions become possible.
At the moment, these corrections are only available for unpolarized
jet production \cite{mi,rom}.
One  expects for the asymmetry $<A>
=\frac{\Delta\sigma^{had}(\mbox{2-jet})}{\sigma^{had}(\mbox{2-jet})}$
that the scale dependence in the individual cross sections
partly cancels in the ratio. 
In fact,  varying the renormalization and factorization scales
between
$ \mu_R^2 = \mu_F^2 = 1/16\;(\sum_j \,k_T^B(j))^2$
and
$ \mu_R^2 = \mu_F^2 = 4\;(\sum_j \,k_T^B(j))^2$
in the LO cross sections
introduces an uncertainty for the ratio $A$ of less than 2 \%,
whereas the uncertainty in the individual cross sections is much larger.

In  conclusion,  the dijets events  from polarized  electron
proton collisions at HERA can provide
a good measurement of the gluon polarization
distribution for $0.002 < x_g < 0.2$, the region
where $x_g\Delta G(x_g)$ is expected
to have a maximum.




\end{document}